# The Power and Pitfalls of Transparent Privacy Policies in Social Networking Service Platforms

**Jana Korunovska, Bernadette Kamleitner and Sarah Spiekermann**

Vienna University for Business and Economics, Vienna, Austria

## ABSTRACT

Users disclose ever-increasing amounts of personal data on Social Network Service platforms (SNS). Unless SNSs' policies are privacy friendly, this leaves them vulnerable to privacy risks because they ignore the privacy policies. Designers and regulators have pushed for shorter, simpler and more prominent privacy policies, however the evidence that transparent policies increase informed consent is lacking. To answer this question, we conducted an online experiment with 214 regular Facebook users asked to join a fictitious SNS. We experimentally manipulated the privacy-friendliness of SNS's policy and varied threats of secondary data use and data visibility. Half of our participants incorrectly recalled even the most formally "perfect" and easy-to-read privacy policies. Mostly, users recalled policies as more privacy friendly than they were. Moreover, participants self-censored their disclosures when aware that visibility threats were present, but were less sensitive to threats of secondary data use. We present design recommendations to increase informed consent.

## INTRODUCTION

Self-disclosure is the main purpose of participation in social networking service (SNS) platforms, and thus, it is in this context that transparency of privacy policies is particularly relevant [1]. SNS users disclose enormous amounts of personal data (PD). For example, as of August 2018, over two billion active Facebook users, 800 million Instagram users, and 330 million Twitter users upload over 450 million photos and 550 million status updates and tweets per day [2-5]. In line with their privacy policies, the SNSs use these data as an economic asset and share it with third parties whereas at the same time their default settings make the users' self-disclosures publically available [6-9]. This often puts users' privacy at risk [1, 10, 11], as is evident from the abundance of recent privacy scandals.

The bottleneck preventing privacy infringements are the SNSs privacy policies. Ideally, privacy policies should make users aware of the risks entailed and instigate privacy-protective behaviors such as avoiding data-invasive SNSs or reducing disclosures in response to the SNSs privacy-friendliness. Despite the general availability of privacy policies, prior research suggests that users hold little awareness of what data are collected and what threats this may entail [12, 13]. A presumed key reason is that policies are difficult to comprehend and inadequate as an information tool [14-17]. As a consequence, users provide consent without being actually informed about what they are consenting to.

In response, scholars and policy makers have called for easily comprehensible, transparent privacy policies [15, 18-23]. The general assumption is that prominent and simple policies will lead to truly informed consent and consequently a more privacy friendly market because informed users will reward privacy-friendly SNSs and punish privacy-invasive SNSs.

Yet, can we be sure that transparent privacy policies will affect PD disclosure behavior as a function of their privacy friendliness? So far we do not have a clear answer to this question. No study to date has investigated how transparency of privacy policy would influence actual disclosure on SNSs depending on whether or not they are privacy-friendly. Even if it is transparent, users may refuse to attend to the policy, miscomprehend or ignore it [14, 24-26].

Here, we design a comprehensive experiment to empirically address three fundamental questions: 1) Will transparent privacy policies on SNSs lead to actual policy comprehension? 2) Will transparent privacy policies prompt users to alter their disclosure behavior in response to how privacy-friendly the policy actually is? And 3) provided there is miscomprehension, will comprehension affect disclosure behavior in response to a policy's privacy friendliness? Though prior studies attest to a lack of comprehension [12, 18, 27], none of them established whether or not this would directly affect self-disclosure.

In addition to addressing these overarching questions, we contribute to the literature by taking a nuanced look at the type of disclosure (quantity and quality) and the type of privacy threat present. Specifically, we focus on two threats that are known to be relevant to users and should thus affect them if made transparent [28-30]. These are PD visibility, i.e. whether others can see the data, and secondary use of PD, i.e. whether SNSs require the right to use the disclosed data beyond what is needed to serve the user. In our experiment, we vary the presence of both threats when manipulating the privacy friendliness of transparent SNSs policies. This adds to earlier insights from research in e-commerce settings [18, 26, 31-36] and makes this one of the first studies to investigate how different types of transparent privacy threats affect disclosure behavior on SNSs.

Eventually the insights pursued here help consolidate two opposing views: one that sees the users as rational agents who, if properly informed, can protect their privacy by altering their disclosure behavior [15, 37] and another that sees users as bounded in their rationality and ultimately manipulable by companies who have high interest in collecting their data [13, 38].

# RELATED WORK

## The Role of Transparent Privacy Policies in Securing Information Privacy

Information privacy is one of the biggest ethical concerns accompanying Internet use today [1, 39, 40] and user data disclosure plays a pivotal role in bringing about these concerns. An abundance of theories has been established to describe and explain self-disclosure online [for reviews see 31, 37, 41, 42]. Based on their underlying assumptions, Barth and de Jong [42] suggest that these theories can be grouped into roughly three approaches. The first approach assumes rational users who decide whether or not to disclose based on a 'privacy-calculus' assessment comparing privacy risks and disclosure benefits. Most prominent in this respect is the APCO model (Antecedents →Privacy concerns → Outcomes), which assumes a relationship between users' awareness of the privacy practices and self-disclosure [37]. The second approach also adopts the idea of a 'privacy calculus' but does not presume a fully rational user. Rather, it assumes that disclosure decisions are bounded by decision-making biases and cognitive limitations. According to this view, users make systematic errors when disclosing personal data and put their own privacy at risk unwillingly. The third approach assumes that users do not perform a privacy calculus and fail to account for privacy risks.

Which of these approaches best reflects reality holds fundamental implications for what effective user-centric privacy protection design ought to look like. Depending on the approach, users are either deemed able to self-protect when provided with all relevant information in a user-friendly way or are deemed incapable of self-protection and need backend design solutions or regulators to protect them. Despite a wealth of research, the empirical evidence on which approach best reflects reality is not straightforward [25, 37] and there are only few attempts to ensure a causal understanding of the phenomenon via experimental research [31, 37]. Until recently, the majority of the community has followed the rational user approach and it has been an exception "for privacy researchers to consider alternative models and explanations outside the APCO model" [32, p. 640].

Policy makers' also largely adopt the rational user approach [32]. For example, in the US, rational users are expected to help the market self-regulate by choosing more privacy-friendly providers [43]. The main role of the regulators, most notably the Federal Trade Commission (FTC), is then to educate the public about online privacy and to recommend to the internet providers good privacy practices (for example, make a clear privacy notice available) and to hold them accountable if they do not adhere to their own policies [8, 44, 45]. In the EU, the recent general data protection regulation (GDPR) likewise assumes that the key to data protection lies in user rights and information [22].

Privacy policies are the dominant information instrument. To allow rational users make informed decisions, policies are expected to provide all necessary information to allow for a privacy calculus. Despite the general availability of privacy policies, ample evidence suggests that users are nonetheless uninformed and fail to self-protect [12, 13]. Note that to be rational users need not be informed. Even ignorance can be rational when the cost of getting the information outweighs the benefit of disclosure [16, 46]. Following this logic, the most discussed reason for uninformed disclosure is policies' inadequacy as an information tool. In fact, privacy policies tend to be long, legalistic, and vague and are not usually written with the user in mind [14-17, 19, 29, 47]. Users also practically never read these hard to understand texts [15, 18, 30, 38, 48].

In response and remaining with the notion of rational users, scholars and policy makers have called for policies to become better information tools, i.e. for them to become easily comprehensible and transparent [15, 18-23]. Specifically, according to the FTC the privacy policies "should be clear and conspicuous, posted in a prominent location, and readily accessible…and also unavoidable" [49, p.7-8], whereas according to the GDPR they should be presented ""in a manner which is clearly distinguishable from the other matters, in an intelligible and easily accessible form, using clear and plain language" [22, p.35].

### The Effects of Transparent Privacy Policies on Policy Comprehension

It is plausible to assume that the majority of users will read and comprehend transparent policies. An eye-tracking study in fact found that users carefully read a privacy policy when presented prominently but at best skimmed it when they had to click on a link in order to access it [30]. Nonetheless, the time people looked at the policy explained only 14 percent of the variance in the knowledge people held about the policy [30]. Consequently, reading a policy appears to be no guarantee for policy comprehension. Internet users have limited time and attention and may not fully comprehend privacy policies for many reasons [16, 25, 50]. By now users are also habituated to accepting terms and conditions without reading them [38, 51]. Even if policies are short and transparent, users may still not engage with and comprehend privacy policies.

Furthermore, transparency is an ambiguous concept [40, 50], not least because it incorporates both prominence (ease of access), and simplicity and conciseness. While simplicity likely increases user comprehension, prominence might have the opposite effect [18, 27, 37]. Some research suggests that a policy's prominence may assure users that there is nothing to hide, so they may skip reading the privacy policies [25, 40]. Even if users read policies, their apparent transparency may affect comprehension. Ample evidence suggests that people see what they expect to see [24]. Consequently, users can "project their privacy expectations onto the privacy notice" [14, p.219]. This may go in two directions. If users interpret transparency as a signal that there is nothing to hide and nothing to worry about, they may miss, misperceive or miscomprehend threats to their privacy even after a cursory reading of privacy policies. In contrast, if users interpret prominent privacy policies as a privacy warning, they may become more concerned [52] and more likely to identify potential threats. Thus, transparency could foster as well as hinder comprehension and informed consent and we assume that some level of miscomprehension will persist even if policies are very simple and transparent. This assumption is a necessary consideration when next hypothesizing how transparency will affect disclosure.

## The Effects of Transparency and Privacy Friendliness on Self-Disclosure

No study to date varied privacy friendliness in the context of prominent policies of SNSs. Related evidence, however, supports policy makers' assumption that information transparency affects users' willingness to disclose PD. Research has found that prominent symbols and signals of privacy (e.g. TRUSTe certificate or the privacy bird indicator) increase privacy-protective behaviors [18, 53, 54]. For example, LaRose and Rifon [34] showed that the presence of a prominent privacy-invasion "warning" decreased respondents intention to disclose PD.

Closer to the matter of manipulating the policy itself, there is additional evidence from the context of ecommerce. Some of this research varied the transparency of privacy policies, other research varied their privacy-friendliness. By and large this evidence appears to suggest that users will react to prominent policies in accordance with the privacy-friendliness of their content. For example, Andrade, Kaltcheva, and Weitz [33, 351] showed that the presence of a prominent short and fully privacy-friendly policy decreased concern for self-disclosure in comparison to a vague notice which stated that the web-site policy is to "respect and protect the privacy" of the users. Likewise, Liu, Marchewka [35] showed that the obvious presence of a privacy-friendly policy increased re-purchase intention (as a proxy for intention to disclose) and, in a survey study, Wirtz, Lwin [55] have shown that perceived friendliness of policies decrease users' intention to upload fabricated or incomplete personal data. In sum, these insights suggest that users will adjust their disclosure behavior in accordance with the privacy-friendliness of a policy. Other studies, however, suggest that this may hold only for users who pay attention to the policy and correctly comprehend it. For example, in a very well simulated ecommerce study by Metzger [36], users disclosed the same amount of PD regardless of whether the policy was very transparent and privacy-friendly, vague and friendly, or entirely lacking. Notably, in the same study about a third of participants reportedly failed to notice the policies when they were present. Similar results are reported by Jensen, Potts [18] who asked participants to compare pairs of e-retailers webpages. In their study only every forth participant actually checked (clicked on) the policies and only then privacy-friendly policies had a bigger effect on the choice of an e-retailer in comparison to privacy-invasive polices. Finally, in a university SNS context, Adjerid, Acquisti [26] showed that privacy notices that warn about higher privacy-risks (faculty staff will be able to view the answers) made students disclose less on sensitive questions (e.g. who was your least favorite professor?) than notices with lower privacy-risks (only students will be able to view the answers). However, these effects were present only when students were undistracted between the presentation of the policy, constraining the effect of the policy notice to ideal circumstances.

Remember that we expect that transparency will not equate to full user comprehension. Although we expect that prominent policies will affect users in accordance with their privacy-friendliness, we do not expect this to hold for users who miscomprehend the policy. Specifically, we expect that comprehension moderates the relationship between privacy-friendliness of a policy and self-disclosure: only those who correctly comprehend the policy can adequately react to its content:

> **H1:** Policy comprehension moderates the relationship between transparent privacy policies and user self-disclosure, thus, users who comprehend privacy-friendly policies will self-disclosure on SNSs more than users who comprehend privacy-invasive policies whereas users who do not comprehend the policy will exhibit similar self-disclosure across different policies.

## The Effects of Specific Privacy Threats on Personal Data Disclosure

Thus far, we have considered privacy friendliness as a unidimensional, dichotomous construct, with a policy being either privacy-friendly or privacy-unfriendly. Though this is in keeping with the common practice in the literature, privacy policies contain several elements that can be more or less privacy-friendly. So far, there is little evidence on how consumers react to different types of threats within a policy. A majority of studies have used scales that combine general privacy concerns (e.g., "How concerned are you about your personal privacy on this website") and specific ones (e.g., "How concerned are you about this website sharing your personal information with other parties?" or "Are you concerned that an email you send may be read by someone else besides the person you sent it to?") [36, 56-58]. Yet, not all of the threats addressed in a policy are deemed relevant by users [28, 29, 59, 60]. For example, users were shown to look significantly longer on the sections "information we collect", "how we use your personal information" and "sharing and disclosure of personal information" than on other sections [30]. In other words, specific privacy threats, rather than a policy's overall level of privacy-friendliness, are likely to drive users' disclosure decisions once they become apparent.

To further advance the current knowledge base, we test for this possibility and focus on two of the most prevalent and dangerous privacy threats in the context of SNSs: secondary data use and public data visibility. Secondary data use refers to a company requesting the right to use user data beyond the primary purpose of data collection. Most of the privacy discussions in politics focus on secondary data use because it holds substantial potential for commercialization and is considered one of the most important causes for concern. A second threat causing concern due to its exploitability is the visibility of PD by an unknown audience. This threat is particularly prevalent in the context of SNS [28, 61, 62].
Users have reason to be concerned about both of these threats. Yet, we propose that they will not react to them equally strongly. Specifically, we propose that users react more strongly to the more easily comprehensible threat of data visibility than to secondary data use. The reason we assume stronger reactions to visibility is twofold. First, visibility is a dominant theme in the privacy discourse specific to SNS. SNS providers themselves foster a focus on visibility when they talk about privacy. For example, under the privacy settings on Facebook the three options are "Who can see my stuff?", "Who can contact me?", and "Who can look me up?". All of these are visibility concerns. Likewise, seven out of eight options in the privacy settings on Twitter are visibility related. SNS users thus have control over the visibility of their data [25, 38, 47] but are rarely invited or able to control secondary data use. Second, an argument has been made that users need to see or at least metaphorically grasp their data to comprehend and protect them [63]. The threat of data visibility is easy to comprehend. People know what it means to know more about another person. They have much less understanding when it comes to the possibilities of secondary data use and the threats this may entail [12, 13].
Some prior evidence supports the proposition that users are more sensitive with regard to visibility than secondary use. For example, Stutzman, Gross [64] analyzed the public Facebook disclosures of more than 5000 students from 2005 to 2011 and showed that over time, the students decreased the amount of PD they shared with the public, but not the PD they shared with Facebook, i.e. they moved the place of disclosure from public walls to private messages which would still fall under secondary use agreements. Similarly, Bazarova and Choi [65] showed that self-disclosure on SNSs varies depending on whether it will be visible to restricted or public audiences, and Das and Kramer [66, p.120] concluded that self-censoring by Facebook users was a function of the "perceived audiences".
Thus, we anticipate that SNS users will be more likely to attend and react to threats of visibility than to threats of secondary data use. This entails two predictions concerning comprehension and disclosure. First and building on the discussed likelihood of miscomprehension despite transparency, we assume a systematic bias in the comprehension of privacy policies:

> **H2:** Transparent privacy policies with secondary data use threats will be miscomprehended more often than transparent policies with visibility threats.

Second, and going beyond the proposed effect on comprehension, we expect an effect of the type of threat on disclosure behavior. We assume that privacy policies highlighting visibility risks will affect user disclosure more than privacy policies highlighting secondary data use. Taking into account that comprehension is a necessary precondition for users to react to any threat we hypothesize:

> **H3:** Among users who comprehend the policy, privacy policies with visibility threats to privacy have stronger effects on self-disclosure than transparent policies with secondary use threats to privacy.

We make no prediction for users miscomprehending the policy because miscomprehension can go into multiple directions.

### A Note on Self-disclosure

Data disclosure can be conceptualized and assessed in multiple ways. To maximize ecological validity, we employ a multitude of disclosure measures. Most previous research has assessed intention to disclose rather than actual disclosure [41, 42]. Even though behavioral intentions predict actual behavior [67], when it comes to privacy, intentions and behaviors are often at odds – a phenomenon known as the privacy paradox [18, 41, 42, 68, 69]. Thus, we complement measures of disclosure intention with actual behavioral proxies. With regard to these proxies, we discriminate between quantity and quality of PD disclosure. Next to quantity, i.e., how much personal information is being disclosed, users can also regulate the quality, authenticity or richness of their disclosures [66, 70]. For example, posting a status on Facebook implies that users have decided to disclose. In addition, users can decide how valuable this disclosure is, that is, how extensive, personally relevant, and self-revealing – in other words how qualitatively rich these data is. We test our hypotheses with regard to all of these conceptualizations of disclosure.

## METHOD

To allow for causal conclusions, we designed an experiment in which we simulated a fictitious SNS (called Coffee Shop) and manipulated the privacy-friendliness of its very transparent privacy policy.

### Stimulus Material

Following FTC and GDPR guidelines on transparency, we focused on the main threats of visibility and secondary data use and designed the Coffee Shop's privacy policy in a simple tabular and additional textual format. This has been argued to best communicate privacy settings [71]. Figure 1 depicts one of these transparent policies. It informs users about who owns, who can see (visibility information) and who can use/sell their data (secondary data use) on Coffee Shop.
We pre-tested the transparency of the policy (n= 60) and confirmed that this policy format was perceived as transparent in terms of simplicity, clarity and easiness of comprehension. On an 11 point semantic differential scale with bipolar adjective pairs (e.g. easy to understand – hard to understand, ranging from -5 to +5) participants rated the language used as very clear (mean=3.47, sd=2.06) and plain (mean=3.27, sd=2.04), and confirmed that it was easy to understand (mean= 2.59, sd=2.64).

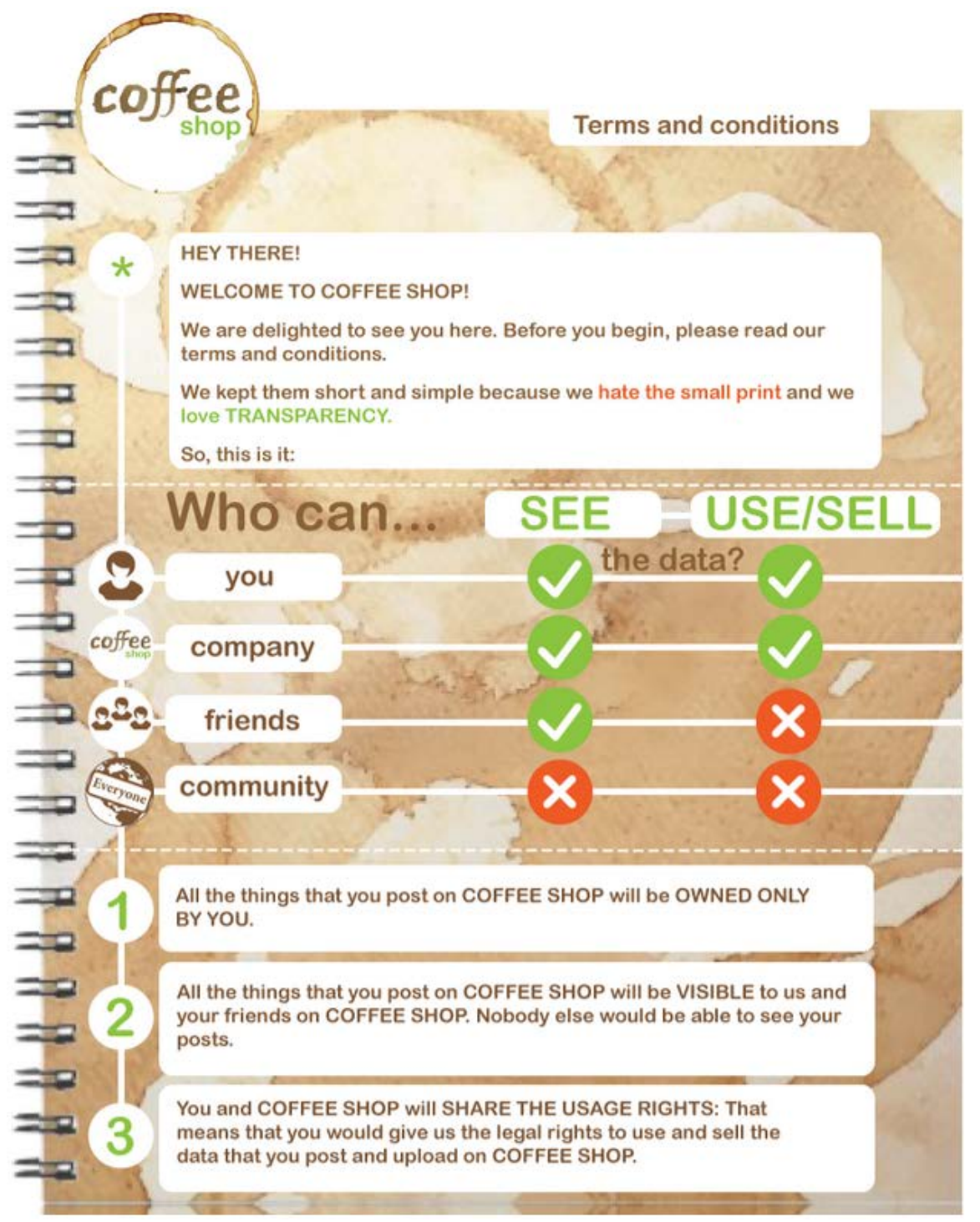

**Figure 1. Example of the privacy policy on Coffee Shop**

## Design, Procedure and Measures

We invited respondents to participate in our study by asking them to "help us learn what makes a good social network". In an initial screening section that goes beyond this study (see Figure 2 for the study flow), we queried respondents about their attitudes and behavior on Facebook as well as their basic demographic data. Next, all respondents landed on an information page with the cover design of several news headers. The main headline read "Coffee Shop trumps Facebook. New social network reaches 2 billion users." We asked respondents to imagine that the SNS "Coffee Shop" had emerged on the market, was already overtaking Facebook in popularity and most of their friends had already joined it. We included this information to preclude Facebook's network effect, a known obstacle to user mobility across SNSs. Additionally, participants learned that this new social network has "a very clear, simple, and TRANSPARENT PRIVACY POLICY" (Figure A, Appendix, see the bottom figure for the stimulus used in the control group). We explicitly positioned Coffee Shop as competing against Facebook on the argument of more transparency about privacy, because this is a known concern on Facebook.

Respondents clicked next to see what Coffee Shop looked like. They landed on a mock up webpage designed to look like a welcome page and actual privacy policy of Coffee Shop. This ensured prominence of the privacy policy and simulated a situation where policies would potentially be unavoidable, following FTC and GDPR guidelines on transparency.

We randomly assigned respondents to one of five conditions consisting of a control group and four different privacy policies. These policies resulted from our 2×2 between-subject design which we employed to test for effects of variations in privacy friendliness. (visibility: entire community can see vs not see the data × secondary data use: SNS can use/sell vs. not use/sell the data) All other aspects of the privacy policy were kept constant across conditions (see Figure 2).

Rather than being presented with a privacy policy participants in the control group were asked to join Coffee Shop without any reference to transparency and also saw no privacy policy. This group, thus, serves as a realistic benchmark control for what happens when privacy policies are hidden and users do not look them up.

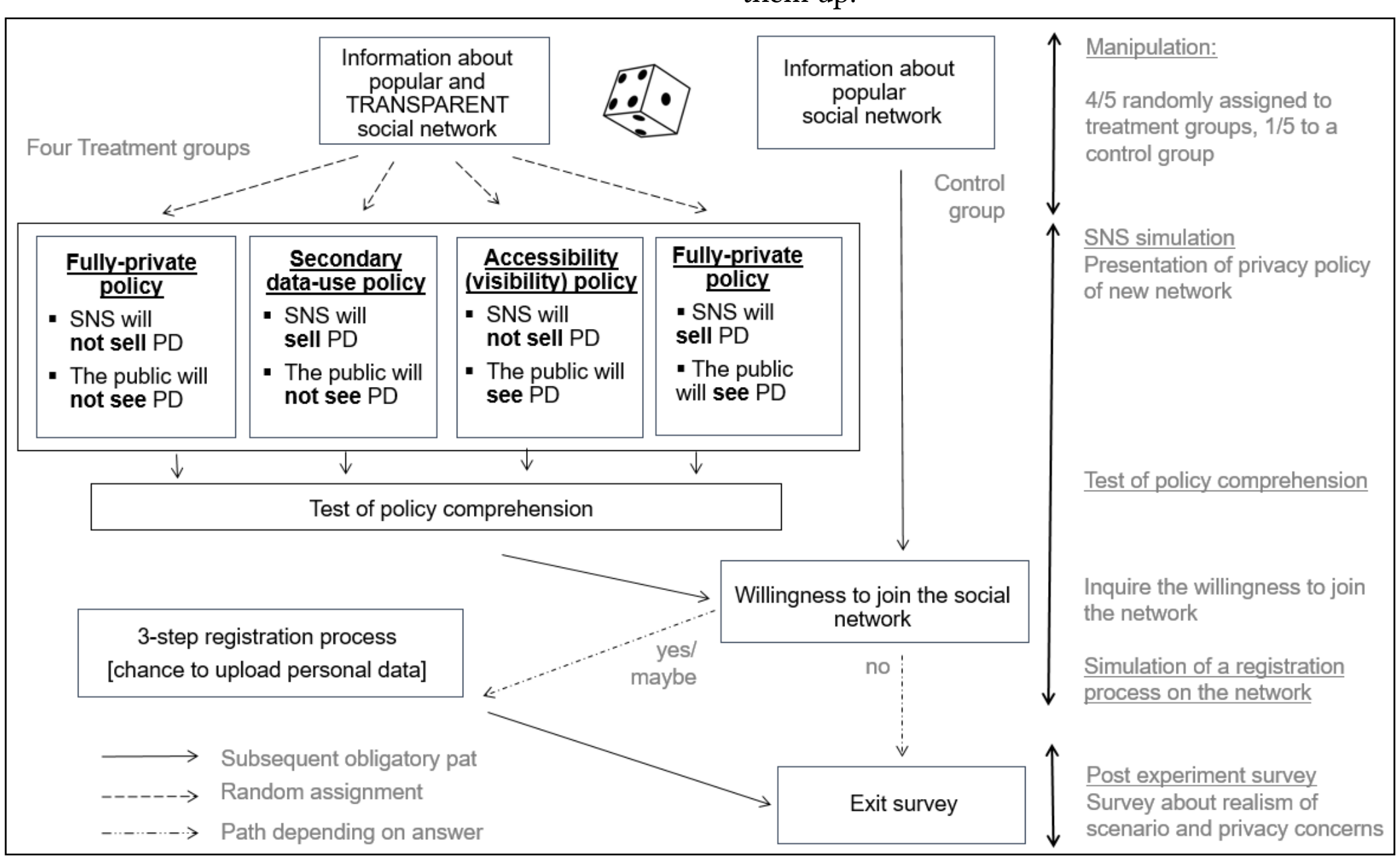

**Figure 2.Study Flow**

### Comprehension via detailed confirmation of policy

Immediately after presenting the privacy policy page to all experimental groups, we subtly checked whether participants had understood the respective policy. In keeping with the cover story, that is, a new platform that authentically values transparency, the social network asked their potential new users to demonstrate that they had read and understood the privacy policy ("It really matters to us that before you sign in you really understand your rights at our shop…"). Users saw the same table from the previous page and were asked to fill in (by ticking the boxes) who can see and use/sell their data on Coffee Shop. Importantly, if unsure, participants had the opportunity to go back and see the actual policy again before replicating the actual ticks, though they could continue even if they got the policy wrong. The correspondence between the ticks made and the actual privacy policy viewed served as a measure of participant comprehension. Not having been exposed to a privacy policy, the control group faced no such comprehension check.

### Disclosure in a simulated 3-step registration process

Next respondents indicated how likely it was that they would actually open an account on Coffee Shop on a 5-point Likert scale ranging from 1="certainly not" to 5="certainly yes". This served as a measure of disclosure intent. All participants who indicated that they had no intention to join at all were thanked and not guided through Coffee Shop any further.

All respondents who indicated that they would be at least somewhat interested to join Coffee Shop next landed on a three-step registration process for the network. The page required their (nick)name and to write in the "about me" section in step one, to upload a profile picture in step two, and to write a first status in step three. In all of these steps the participants could choose not to disclose any of their personal data To simulate the potential transfer from Facebook and to comply with the data portability principle in the GDPR, respondents could also click to indicate that they would import their name and profile picture from Facebook.

To assess the quantity of self-disclosure we summed up disclosure decisions across the three steps. We coded every disclosed item (name, about me, photo, and status) with one when provided and zero when not provided. The resulting disclosure index could take on values between zero and four.

In addition, we assessed a proxy for the richness of the self-disclosures by coding the content of the status posts, which was the only disclosure item that could vary in richness. We coded the status messages with zero if they respondents did not write a status, with one if they wrote only "Hi", "Hello" or similar, and with two if they wrote more personal statuses (e.g., "I'm up for something new, interesting, and exciting :-)") thus making richer disclosures.

After the registration process, respondents were redirected back to the survey where we asked them how realistic they found the SNS transfer scenario to be; from 1="very unrealistic" to 5="very realistic". In order to identify respondents who were not paying attention to the survey, we incorporated two attention-testing questions, one before and one after the experimental manipulation. We placed the questions in-between scales so that they would be missed by people who simply clicked through without paying attention (e.g., "please click 'rarely' if you are still paying attention"). Participants failing to pass these checks were screened out.

## Sample

A professional market research company administered the study to their internal online panel of over 30.000 respondents. This panel is a nationally representative sample of Internet users in Austria. The respondents were randomly drawn from this sample to ensure representativeness. All invited panel users were first asked whether they hold a Facebook account and could thus believably simulate a transfer to a new account. If they did not, they were screened out automatically. The final sample consisted of 214 Facebook users (55 percent female, mean age=40 years; sd=4.5 years) who completed all questions and passed the attention checks.

## RESULTS

Groups were structurally equivalent in terms of their demographic makeup. Namely, all groups were of similar age (f(4, 208)=0.52, n.s.), gender (f(4, 207)=0.32, n.s.) and had similar levels of education ($\chi^2(1)=0.72$, n.s). Also, all experimental groups found the scenario to be equally realistic (m=3.09, se=1.23, f(3, 209)=1.28, n.s,). Presumable due to the lack of a clear unique selling point of Coffee Shop over Facebook (i.e. transparency), the control group found the transfer scenario to be less realistic than the experimental groups (m = 2.50, se = 1.26, t(176)=2.51, p<.05) .

## Comprehension of Transparent Privacy Policies

The primary assumption behind the proposition of privacy policies is that they users will attend to them, comprehend them and eventually provide informed consent. As outlined above, we anticipate that this may be too optimistic a view. We thus first looked at the effect of transparent policies on policy comprehension. To do so we checked whether respondents correctly replicated the rights and threats that the privacy policy comprised, by ticking the exact same boxes.

Comparing the simple and unavoidable policies participants had seen to those they recalled only seconds later yielded a picture of considerable miscomprehension. Overall, only 53 percent of the respondents were able to correctly recall all the respective policy's content, with a 95 percent confidence interval between 46 and 61 percent

In a next step we looked to test H2 and see whether secondary use threats were more likely to be miscomprehended than visibility threats. To do that, we ran a logistic regression with policy comprehension as the outcome variable and the specific policy elements (secondary data use threat, visibility threats) as predictor variables. Results show that those participants who were exposed to the threat of secondary data use were more than twice as likely to miscomprehend this information in comparison to those participants who were not exposed to this threat (β=0.81, se=.32, or =2.25, p<.05). Exposure to the visibility threat was not associated with the likelihood of miscomprehension (β=-0.48, se=.32, OR=.62, n.s).

**Table 1. Privacy Notice Comprehension: What Respondents Saw and What They Recalled**

| | | Respondent recalled: | | | Respondent recalled: | |
|---|---|---|---|---|---|---|
| | | **Company will not sell PD** | **Company will use/sell PD** | | **The ublic will not see PD** | **The public will see the PD** |
| **Privacy notice showed:** | **Company will not use/sell PD** | 93% | **7%[a]** | **Public will not see PD** | 72% | **28%[c]** |
| | **Company will use/sell PD** | **35%[b]** | 65% | **The public will see the PD** | **24%[d]** | 76% |
| | N | 90 | 85 | | 97 | 78 |
| Note. In bold are respondents who did not comprehend the privacy notice: a. overestimated secondary threats, b. underestimated secondary threats, c. overestimated visibility threats, d. underestimated visibility threats | | | | | | |

In a next step we looked to test H2 and see whether secondary use threats were more likely to be miscomprehended than visibility threats. To do that, we ran a logistic regression with policy comprehension as the outcome variable and the specific policy elements (secondary data use threat, visibility threats) as predictor variables. Results show that those participants who were exposed to the threat of secondary data use were more than twice as likely to miscomprehend this information in comparison to those participants who were not exposed to this threat ($\beta$=0.81, se=.32, or =2.25, p<.05). Exposure to the visibility threat was not associated with the likelihood of miscomprehension ($\beta$=-0.48, se=.32, OR=.62, n.s).

As predicted in H2 information on secondary data use was associated with policy miscomprehension versus visibility threats were not. Notably, participants were more likely to miscomprehend the policy in an unsafe direction, i.e. miscomprehension was associated with a wrong sense of security (Table 1). Participants' most frequent error was to think that the SNS had no rights to share and sell their data (i.e., secondary data use) when in fact it had. One in four participants also erred with regard to who can see their data, but this went both in the safe as well as in the unsafe direction.

## Transparent Privacy Policies and Self-disclosure

The next question we addressed was whether a transparent policy's privacy friendliness would affect disclosure specifically among those who comprehended it (H1) and whether this effect would depend on the type of threat present (H3). For this purpose, we established a new variable which divided respondents into those who correctly recalled the policy and those who did not. We used this variable and the manipulation of threat of visibility and threat of secondary data use to predict disclosure. We ran ANOVAs on the willingness to join the SNS and the number of uploaded items and conducted ordinal regression and $Chi^2$ tests for the richness of disclosed PD.

*Willingness to join the SNS.* A three-way ANOVA on users' willingness to join the new SNS revealed a significant main effect of the presence of a visibility threat (f(1, 167)=3.98, p<.05); a significant main effect of policy comprehension (f(1, 167)=14.82, p<.001); and a significant three-way interaction between visibility threat, secondary data use threat, and policy comprehension (f(1, 167)=4.42, p<.05). As evidenced by the respective main effect, the willingness to join was higher among those who comprehended the policy and by those who did not face a visibility threat. To facilitate the interpretation of the three-way interaction, Figure 3 shows the pattern of results among those who (a) did not and (b) those who did comprehend the privacy policy. According to H1 we expect that the policy content will only affect those who comprehend it. According to H3 we expect that this holds in particular for the visibility threat.

In line with these predictions, we find no difference in willingness to join among those who did not comprehend the policy (f(2,83)=.11, n.s). Among those who did comprehend the policy we find the expected effect of visibility threats (f(1,92)=6.95, p<.01). If visibility threats were correctly reported as present, participants were less likely to join. The presence of secondary data use threats also affected joining intentions, but to lesser extent (f(1,92)=2.88, p<.05).

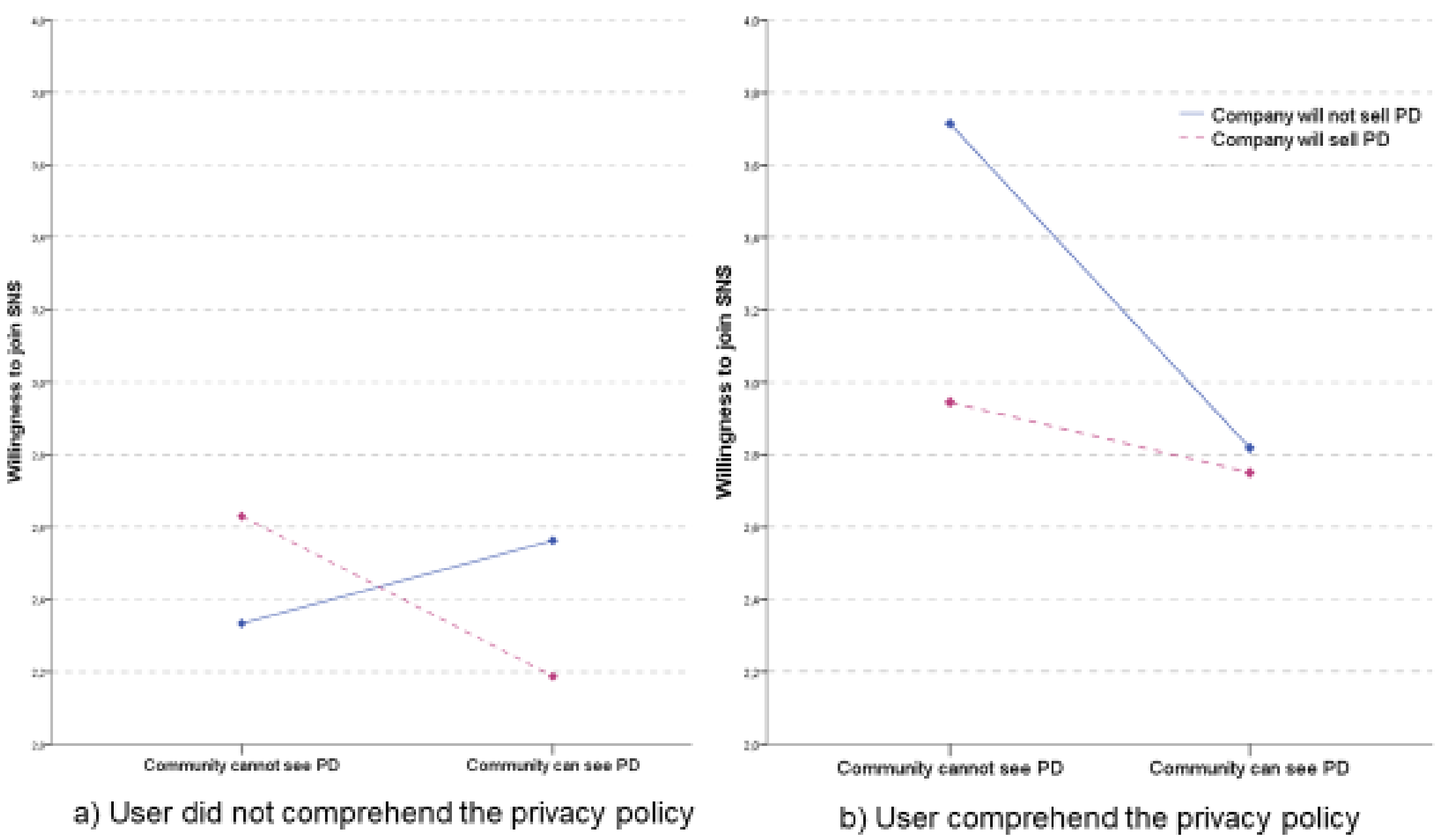

**Figure 3. Willingness to join the SNS based on policy comprehension and type of privacy threat**

*Number of uploaded PD on the SNS.* A three-way ANOVA on the number of uploaded items showed only a significant two-way interaction between visibility threats and policy comprehension (f (1, 136) = 4.19, p < .05) which is illustrated in Figure 4. In line with H1 the policies content only affected the number of PD respondents decided to disclose if participants comprehended this content. In line with H3, only one content, visibility threats, did yield this effect. Respondents disclosed less, if others could see their disclosed content. Although the pattern emerging resembles that for visibility threats, the threat of secondary data use did not significantly affect disclosure

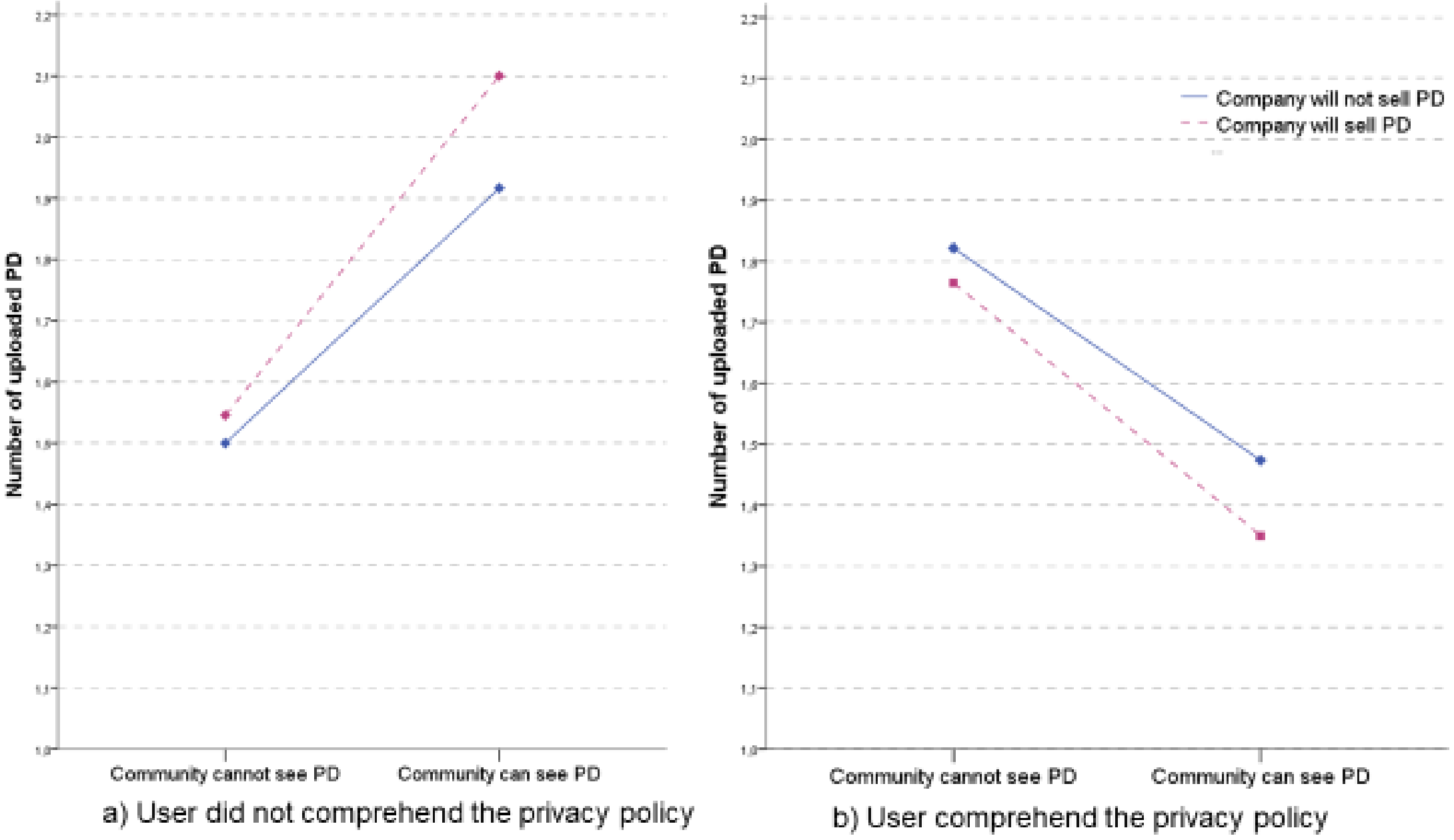

**Figure 4 Number of uploaded PD depending on policy comprehension and privacy threat**

*Richness (quality) of disclosed information.* To test how comprehension affects the relationship between transparent policy and richness of the self-disclosure, we ran two ordinal regressions, one for those who miscomprehended and one for those who comprehended the policies.

The results showed that among those who miscomprehended the policy, neither threat had influence on the richness of the disclosed status. However among those who comprehended the policy, we found significant difference similar to the results for quantity of disclosure.

**Table 2. Status Richness (Quality) Depending on the Threats Present in the Notice and on Notice Comprehension**

| Status richness (quality) | Users who comprehended a privacy notice that stated: | | | | Control group |
|---|---|---|---|---|---|
| | Company can sell (Secondary data use threat) | | Community can see (visibility threat) | | |
| | No | Yes | No | Yes | |
| Did not write a status | 30% | 44% | 24% | 49% | 32% |
| Wrote just Hi or Hello | 26% | 24% | 27% | 23% | 29% |
| Wrote a more personal status | 45% | 31% | **49%*** | **28%*** | 38% |
| N | 47 | 45 | 45 | 47 | 34 |
| Note: * $\chi^2(1) = 4.39$, $p < 0.05$; Percentages are subject to a rounding error | | | | | |

Namely, respondents who saw and comprehended a SNS privacy policy that entailed a visibility threat wrote less personal statuses than respondents who correctly comprehended that their data would not be visible ($\chi^2(1) = 4.39$, $p < 0.05$). In contrast, the notification that the SNS will use their data commercially did not significantly affect richness of disclosure, although the direction was in the expected direction (see Table 2).This final result also supports the anticipated effect of comprehension and type of threat hypothesized in H1 and H3.

## DISCUSSION

Informed consent is the pinnacle of user self-protection and policy makers and scholar have called for SNS platforms to provide users with simple and transparent privacy policies. But will the change of privacy policy design yield informed consent? To answer this question, scholars have called for more experimental studies conducted with representative samples and actual disclosure measures [31, 32, 41]. By conducting an experimental simulation of the process of joining a new SNS, we have answered to these calls. We find that transparent privacy policies are insufficient to ensure informed consent. Despite placing the policy prominently and even unavoidably, despite focusing on threats we know users care about, despite using a policy design with proven simplicity and despite forcing participants to engage with the policy by asking them to replicate it, half of our representative population sample of Facebook users did not correctly comprehend the policy. This is even more remarkable if we consider that the sample consisted only of participants who had passed two attention checks and that respondents could go back and double check the policy. Our sample of SNS users was thus unable to provide informed consent and- to a large extent- did not disclose data as a function of the risks the policy entails. This puts into question whether truly informed consent is at all attainable and is in direct contrast with the rational user hypothesis [72].

Of particular concern is that miscomprehension was systematically in favor of the SNS rather than the user. The most frequent recall error was that respondents thought that the company will not allow secondary data use, even though it transparently announced that it will. It could be that common practices by popular SNS providers which draw attention to visibility rather than secondary data use have habituated users in disregarding these threats [25, 51, 73]. A potential other explanation is that transparency signals trust and thus leads users to misperceive the content of the policy in privacy-unsafe directions [14]. More research is needed to unravel the full reasons behind users' miscomprehension and its effect on self-disclosure.

Another key finding is that users do react to visibility threats once these are salient. Users who were aware that their data will not be publicly accessible disclosed more and richer PD, suggesting that salient visibility threats lead users to self-censor their disclosures. The secondary data use threats play no such role and do not affect disclosure. This is noteworthy because secondary data use may be an even bigger actual threat. The right to secondary data use may pertain to all data that users provide. This entails log in data etc. which users may not even be conscious of. As users learn that visibility threats can lead to secondary data use threats (as was the case with the Facebook-Cambridge Analytica data scandal) and as they gain more right to control over secondary data use through the GDPR, it will be interesting to see if the influence of the two threats on self-disclosure changes.

### Design and Policy Implications

The study at hand shows that we cannot expect that transparent privacy policies are a stand-alone solution for gaining users'

informed consent. This is why efforts to make privacy policies machine-readable, either by natural language processing efforts [74] or with the help of privacy protocols [75], are one necessary addition. Stricter regulations and potentially also user education are other options that can mitigate this problem [55, 71]. However, given that the information was already presented in a very simple and prominent manner it is doubtful whether user education may really solve the issue. Users appear to have become accustomed to all or nothing offers in exchange for (among other things) privacy-invasion [38]. Thus, designing comprehension checks in the privacy policies (such as proof of direct user engagement in a captcha-like proof of comprehension and not just tick of a box) seems essential.

Our results also hold implications for what type of threats may be in particular need of designers and policy makers' attention. Users seem to care more about visibility than secondary data use even though secondary data use can potentially be more harmful to them [76]. Our results show that users (a) systematically underestimate the risk of secondary data use when engaging with a privacy notice and that (b) even if they recognize this risk they still behaviorally disregard it. Therefore, our results primarily suggest the need for designers to highlight the secondary data use threats. In addition, enhanced user control over secondary data use might help mitigate the current failure of self-protection. Though most SNSs have public access as their default setting, almost all of them allow users to restrict access and visibility to their personal data [47] but do not allow similar controls when it comes to secondary data use. If those solutions are not embedded in the design, policy makers will need to protect the users with regard to the right of use to their data where their own consciousness of consequences appears limited.

In sum, if we are to have informed SNS users, we need more than just simple and transparent privacy notices. At this stage of progression into the digital age a lot of users will not read or comprehend them, and even when they do, they will not punish privacy-invasive services, and at best reward privacy-friendly services with more disclosure. Also they will systematically underestimate the risk of secondary data use. We may have to drop the rational user hypothesis and remove the burden of protecting their privacy against hard to understand, powerful players from the shoulder of users altogether.

## Limitations and Future Research

We are aware that our study has limitations that should be addressed in future research. Our sample was representative only of Internet users in a European country. Given that privacy norms and expectations have been found to be culturally dependent [77], it would also be interesting to establish the extent to which the results generalize to other cultural contexts in which consumers may be more or less privacy aware.

Furthermore, social network users have been shown to have higher risk taking attitudes [78], and it is thus possible that users' propensity to disclose in the face of transparent privacy-invasiveness is particularly pronounced in the context of SNSs and cannot be generalized further. Future research will be needed to look into the extent to which our insights generalize to other contexts featuring privacy policies such as ecommerce, entertainment, and information applications. It appears possible that in these contexts where visibility threats play less of a role, users are more sensitive to the threats of secondary data use.

It would also be interesting to extend the inquiry to additional privacy threats. We specifically considered two privacy threats that are known to matter in the context of SNS [28, 29, 59, 60] and thus the privacy policy was oversimplified. It will take a lot of effort by the community and the regulators to overcome the trade-off between simple and short and also comprehensive privacy policies that also include other relevant information like all the type of PD, type of processing, storage, security, revoke of consent etc.

## Appendix

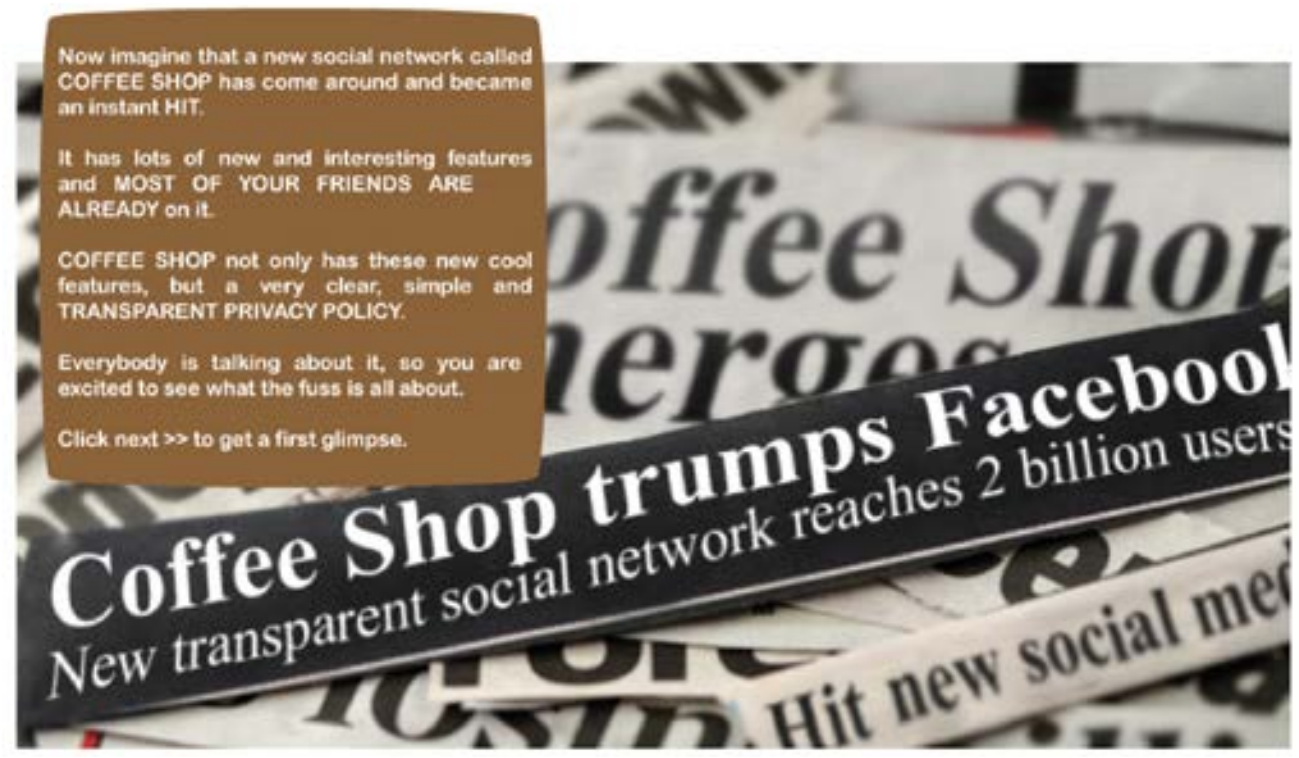

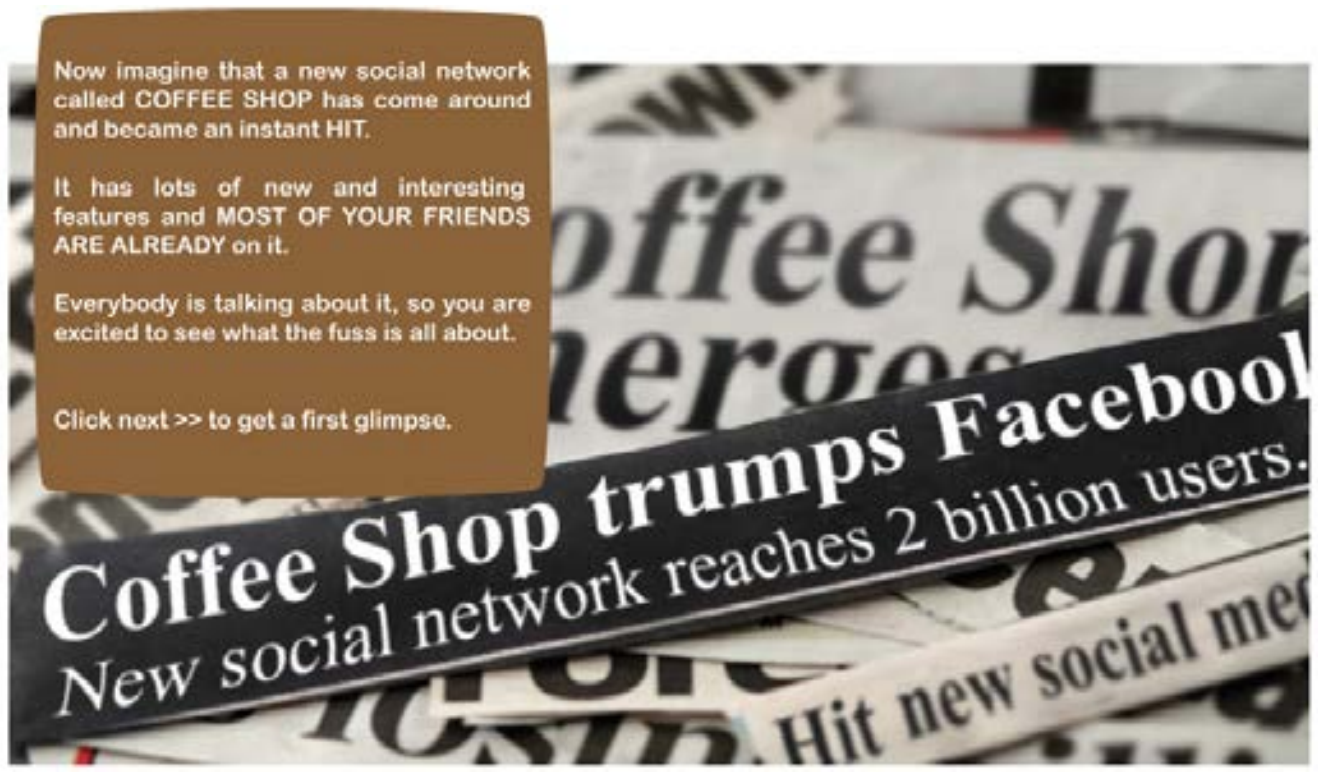

**Figure A. The Introduction to the Hypothetical New OSN "Coffee Shop" Scenario for (above) the Experimental Groups and (below) the Control Group**